\documentclass{llncs}
\usepackage{url}
\urldef{\mailsa}\path|{alfred.hofmann, brigitte.apfel, ursula.barth, christine.guenther,|
\urldef{\mailsb}\path|ingrid.haas, frank.holzwarth, anna.kramer, leonie.kunz, nicole.sator,|
\urldef{\mailsc}\path|erika.siebert-cole, peter.strasser, lncs}@springer.com|

\usepackage{amsmath}
\usepackage{amssymb}
\usepackage{graphics,color}
\usepackage{graphicx}
\usepackage{indentfirst}
\usepackage{algorithmic}
\usepackage{algorithm}
\usepackage{extarrows}
\usepackage{subfigure}

\newcommand{\MPC}{{\ensuremath{\sf MPC}}}
\newcommand{\MCC}{{\ensuremath{\sf MCC}}}

\newcommand{\setup}{{\text{\sc Setup}}}
\newcommand{\keygen}{{\text{\sc Keygen}}}
\newcommand{\proxykeygen}{{\text{\sc ProxyKeygen}}}
\newcommand{\thkeygen}{{\text{\sc ThresholdKeygen}}}
\newcommand{\enc}{{\text{\sc Enc}}}
\newcommand{\proxyenc}{{\text{\sc ProxyEnc}}}
\newcommand{\homevaluation}{{\text{\sc HomoEval}}}
\newcommand{\dec}{{\text{\sc Decrypt}}}
\newcommand{\thdec}{{\text{\sc ThresholdDec}}}
\newcommand{\parameter}{{\ensuremath{\sf param}}}
\newcommand{\ssk}{{\text{\sc ssk}}}
\newcommand{\spk}{{\text{\sc spk}}}
\newcommand{\rk}{{\text{\sc rk}}}
\newcommand{\outp}{{\text{\sc output}}}

\newcommand{\Z}{\mathbb Z}
\newcommand{\pk}{{\text{\sc pk}}}
\newcommand{\sk}{{\text{\sc sk}}}

\newcommand{\ignore}[1]{}
\begin{document}
\title{Multiparty Cloud Computation}


\author{Qingji Zheng \inst{1}
\and Xinwen Zhang  \inst{2}
}

\institute{University of Texas at San Antonio, TX, USA \\
qzheng@cs.utsa.edu
 \and
Huawei Research Center, Santa Clara, CA, USA \\
xinwenzhang@gmail.com
}

\maketitle
\begin{abstract}
With the increasing popularity of the cloud,  clients oursource their data to clouds in order to take advantage of  unlimited virtualized storage space and the low management cost. Such trend prompts  the privately oursourcing computation, called \emph{ multiparty cloud computation} (\MCC):  Given $k$ clients storing their data in the cloud, how can they perform the joint functionality  by contributing their private data as inputs, and making use of cloud's powerful computation capability. Namely,  the clients wish to oursource computation to the cloud together with their private data stored in the cloud, which naturally happens when the computation is involved with large datasets, e.g., to analyze malicious URLs. We note that the \MCC\ problem  is different from  widely considered concepts, e.g., secure multiparty computation  and  multiparty computation with server aid.

To address this problem, we introduce the notion of \emph{homomorphic threshold proxy re-encryption} schemes, which are encryption schemes that enjoy three promising properties: proxy re-encryption -- transforming encrypted data of one user to encrypted data of target user, threshold decryption -- decrypting encrypted data by combining secret key shares  obtained by a set of users, and homomorphic computation -- evaluating functions on the encrypted data.
To demonstrate the feasibility of the proposed approach, we present an encryption scheme  which allows  anyone to compute arbitrary many additions and at most one multiplications.

\end{abstract}

\section{Introduction}
\noindent
The concept of cloud computing has been widely accepted by individuals and enterprises. It is getting more and more popular because ideally it can provide unlimited computation capability and storage space by virtualizing vast physical computer resources and integrating them together. Due to these advantages as well as the low management cost, individuals and enterprises  have been taking actions to outsource to clouds their data and hopefully outsource computation,  which is still a challenging task which yet has not been well resolved in the literature.

In the present paper, we consider the problem that clients (individuals or enterprises) wish to oursource computation to the cloud together with their private data stored in the cloud. More specifically, we consider the problem formulated as follows:

\medskip
\noindent \textbf{Multiparity Cloud Computation(\MCC)~~}Consider that $k$ clients, $p_1, \cdots, p_k$, store their data $x_1, \cdots, x_k$ in clouds in an encrypted form, they wish to cooperate together in order to efficiently and securely  compute the function $f(x_1, \cdots, x_k)$ by utilizing the computation capability of clouds.

\medskip
\noindent Here the terms of ``efficiently" and ``securely" pose some intuitive requirements towards solving \MCC\ problem:
\begin{enumerate}
\item The communication overhead between the clients and the cloud should be minimized for the purpose of efficiency. It will rule out some trivial solutions  where the clients download their data from the cloud, decrypt them to obtain the original data, and then adopt secure multiparty computation protocols with cloud aid. Since downloading data from cloud will pose heavy communication overhead, it dispels the benefit of the cloud. In addition, the evaluation of functionalities should be performed in the cloud in order to take advantage of cloud computation capability.
\item The data privacy should be preserved for the purpose of security, which has three fold: (i) The data stored in the cloud should be kept privacy, namely the data should be encrypted before outsourced to the cloud. (ii) The evaluation result should be kept private from the cloud, (iii) each client can not learn anything other the result of the evaluation $f$ and the information revealed by that.
\end{enumerate}

While \MCC\ problem is quite similar to multiparty computation (\MPC) problem ~\cite{Yao:1982} at first glance, indeed it is different from that. As in the multiparty computation problem, the inputs of the evaluation reside at the client side. Moreover, the evaluation will be performed by the clients themselves, rather than by the cloud in the case of \MCC. Those two distinctions make the \MCC\ more difficult than \MPC\ because of the efficiency demand and the privacy issue stated as above. More information about \MPC\ can be referred to \cite{Oded:2009}.

Another related problem is secure multiparty computation with server aid, which was recently introduced in~\cite{Kamara:2011}.  The secure multiparty computation with server aid can be regarded as a bridge between the gap \MCC\ and \MPC\ because the inputs of the evaluation are still stored locally similar to \MPC\, but the evaluation will be performed in the cloud analogous to \MCC.

\ignore{
that can be dated back to Yao~\cite{Yao:1982} and has been widely and extensively studied by a large number of works (cf.~\cite{Oded:2009} and its references):  $k$ parties, $p_1, \cdots, p_k$, with their respective inputs $x_1, \cdots, x_k$, cooperate to compute a function $f(x_1, \cdots, x_k)$. However, \MCC\ holds some subtle but essential difference from secure multiparty computation: data (inputs) are hold in individual client sides secure multiparty computation, while they are resided in the cloud in \MCC; the evaluation of the joint function is executed by individual clients in secure multiparty computation, while it is in the cloud in \MCC.
With these differences, \MCC\ protocol poses some unique requirements beyond that in secure multiparty computation, which usually only desires the first above.
}

\ignore{
The straightforward but trivial solution to \MCC\ is that $k$ parties download their owned data from the cloud respectively, and then execute a secure multiparty computation protocol \cite{cryptoeprint:2000:055,cryptoeprint:2011:535} or a secure server-aided multiparty computation protocol \cite{Kamara:2011}. While this solution can correctly and securely compute the function $f$, it is not efficient in the sense of computation and communication: the clients have to download the data from the cloud and then decrypt them, which obviates the benefits offered by cloud computing.
}

\medskip
\noindent{\bf Applications~~}Among many others, here we present two applications about multiparty cloud computation.

\begin{enumerate}

\item Malicious URLs Detection: In order to automatically detect malicious URLs, the researcher may propose a complicated model which involved many features distilled from large dataset, e.g. malicious URL samples and web page pointed by the URLs. To make the model more accurate, the detection model should be trained with a mass of data from many anti-virus companies. However, these companies store the data in the cloud for the sake of economy  and are unwilling to share the data directly to the researchers. Therefore, the researcher has to outsource the evaluation (detection model) to the cloud, so that the cloud could perform the evaluation by taking as inputs  the data from companies.

\item Healthcare Information Query: Healthcare service providers (hospitals) maintain their patient records and now would outsource these databases to the cloud. They may encrypt the database in order to preserve patient privacy in order to comply with some regulations. Some research group tries to estimate the trend of certain disease by analyzing the symptoms from a large number of patients. The hospitals are reluctant to share their databases with the research group, but only allow the group to perform evaluation on top of their data. Hence, the research group has to describe their estimation model and let the cloud perform the model instead.
\end{enumerate}

\section{Model Formulation}

\noindent
{\bf System Model~~} We consider three types of entities in our model: a third party, a service provider providing service in the cloud, and many clients making use of cloud service. Typically, the service provider may not be the cloud vendor, although we refer them as the cloud in short here. The third party is independent from the clients and the cloud, and  responsible for key management when considering the cryptographic primitives.

The clients not only outsource their data to the cloud, but also outsource the computation functions, which can be any models to analyze or estimate the data. The cloud hosts the data owned by clients in a isolated manner -- individual clients' data are separated from each other, and the cloud provides certain level of reliability, e.g., satisfying some SLA agreed with clients. The cloud will evaluate the functions for the client and eventually the clients learn the result but the cloud does not learn anything from that. We further assume there exists  a secure and authenticated communication channel between any two entities.

\medskip
\noindent \textbf{Adversarial Model~~}
The security threat originates from the misbehavior of the clients and the cloud. We consider a computationally bounded adversary model -- semi-honest but curious model, which specifies the behaviors of the clients and the cloud. Specifically:
\begin{enumerate}
    \item The clients and the cloud execute the protocol's specification exactly;
    \item The cloud provides reliable storage service, namely it does not modify or destroy the stored data;
    \item The inputs of the function are provided appropriately. Some techniques, e.g keyword search, can be adopted to facilitate the cloud to prepare dataset for the function.
    \item The cloud is curious and makes great effort to infer something from the execution;
    \item While the client may be reluctant to leak any information related to its own data stored in the cloud, it is desire to learn information from other clients's dataset.
\end{enumerate}

We also emphasize that the third party is fully trusted. The trusted third party will be responsible for issuing keys, and managing key distribution as needed.

\section{Homomorphic Threshold Proxy Re-Encryption Scheme}
\label{sec:definition}

\noindent In order to preserve privacy,  the clients will encrypt their data when they outsource it to the cloud. However, the encrypted form of data greatly impedes the utilization due to its randomness.
Many efforts have been done for the purpose of data usage but without undermining the data privacy. Homomorphic encryption has been one of critical techniques to achieve this objective and can be found in volumes of research work~\cite{Gentry:2009,Gentry:2011,cryptoeprint:2011:000,cryptoeprint:2011:344,cryptoeprint:2011:441}.  However, simply adopting the homomorphic encryption does not work  in \MCC\ because of the fact that homomorphic encryption scheme only could perform homomorphism evaluation in the case of  ciphertexts under the same public/private keys. In \MCC\ the inputs of the functions are from multiple clients with their own public/private key, which prohibits adopting  homomorphic encryption schemes directly. Hence, this inspires us to introduce the proxy re-encryption capability into the homomorphic encryption schemes.

In practice, we propose the \emph{homomorphic threshold proxy re-encryption} schemes with desirable properties:
\begin{itemize}
\item Homomorphism: Given two ciphertexts $c_1$ and $c_2$ on plaintexts $m_1$ and $m_2$ respectively, one can obtain the ciphertext on the plaintext $m_1 + m_2$ and/or $m_1 \cdot m_2$  by evaluating $c_1$ and $c_2$ without decrypting ciphertexts.
\item Proxy re-encryption: Given a proxy re-encryption key, the proxy can transform a ciphertext of one user to a ciphertext of the target user.
\item Threshold decryption: By dividing the private key into several pieces of secret shares, all clients can work together to decrypt the ciphertext -- the output of the function.
\end{itemize}

\subsection{Scheme Definition}

\begin{definition} (homomorphic threshold proxy re-encryption)
A homomorphic threshold proxy re-encryption scheme, denoted by $\Lambda$, consists of the polynomial-time algorithms as follows:
\begin{description}
\item[\setup($1^\ell$):] Given security parameter $1^\ell$, this algorithm outputs the global parameter \parameter, which includes the specification of message space, plaintext space, and  ciphertext space. We assume that  \parameter\ is implicitly included as input in the following algorithms.

\item[\keygen:] This algorithm generates a public/private key pair $(\pk_i,\sk_i)$ for client $i$.

\item[\thkeygen($k$):] Here $k$ is the expected number of secret shares. This algorithm generates a pair of  public/private key, and divides the private key into $k$ shares, with which  $k$ clients together can decrypt a ciphertext encrypted with the corresponding public key. We denote the public/private keys denoted as $(\spk, \ssk)$, where $\ssk = \{\ssk_1,\cdots,\ssk_k\}$. We name $(\spk, \ssk)$ will be the target public/private key.

\item[\proxykeygen($\sk_i,\spk$):] This algorithm allows client $i$ to generate pa roxy re-encryption key $\rk_{i}$  by taking as inputs its private key $\sk_i$ and the target public key $\spk$, such that the proxy can re-encrypt a ciphertext under $\pk_i$ to a ciphertext under $\spk$.

\item[\enc($M,\pk_i$):] Given the message $M$ and public key $\pk_i$, this algorithm encrypts $M$ and outputs ciphertext $C$.

\item[\proxyenc($C, \rk_{i}$):] Given ciphertext $C$ under public key $\pk_i$ and a re-encryption key $\rk_{i}$,  this algorithm will output ciphertext $C'$ under public key $\spk$ via re-encrypting $C$.

\item[\homevaluation($\{C'_1,\ldots,C'_k\}, \spk, f$):] Given a set of ciphertext ${C'_1,\ldots,C'_k}$ under the public key $\spk$, corresponding to the messages $m_1,\ldots,m_k$, this algorithm generates ciphertext $C'_h$, such that $C'_h = \enc(f(m_1,\ldots,m_k),\spk)$.

\item [\dec($C, \sk_i$):] Given ciphertext $C$ under public key $\pk_i$, this algorithm decrypts the message $m$ from $C$ with private key $\sk_i$.

\item [\thdec($C'_h, \ssk= \{\ssk_1,\ldots,\ssk_k\}$):] Given ciphertext $C'_h$ under public key $\spk$, this algorithm decrypts the message $m$ from $C'_h$ with the cooperation of $k$ clients holding secret share $\ssk_i, 1\leq i\leq k$, respectively.
\end{description}
\end{definition}

\subsection{\MCC\ Protocol}

\noindent
With the the homomorphic threshold proxy re-encryption scheme $\Lambda$ defined as above, we can construct protocol to solve \MCC\ problem.  Let $\{(P_1,M_1),\ldots,(P_k,M_k)\}$ be the set of pairs of clients and its own data involved in the \MCC\ problem. Note that $M_i$ will be outsourced to the cloud in an encrypted form.  Note that we assume all clients and the trusted third party share the common parameter generated by $\Lambda.\setup.$

\smallskip {\bf \noindent Setup Phase:}
\begin{itemize}
\item  The trusted third party invokes $\Lambda.\setup$ and initializes the public parameter of $\Lambda$, which will be shared by all clients.

\item Client $i$ invokes $\Lambda.\keygen$ and generates public/private key pair ($\pk_i,\sk_i$)

\item Client $i$ encrypts its data $M_i$ with the public key $\pk_i$, obtains $C_i$, and then outsources $C_i$ to the cloud.
\end{itemize}

{\bf \noindent Preparation Phase:}
\begin{itemize}
\item The trusted third party invokes $\Lambda.\thkeygen$ and obtains a pair of public/private key ($\spk,\{\ssk_1$, $\cdots, \ssk_k\}$). It publishes the public key $\spk$ and
    distributes the share of private key $ssk_i$ to the  client $i$ through a secure channel, assuming there has $k$ clients contributing their data.

\item Client $i (1\leq i \leq k)$  invokes $\Lambda.\proxykeygen$ by taking as inputs $\sk_i$ and $\spk$, and generates a proxy re-key $\rk_{i}$. Then client $i$ sends $\rk_i$ to the cloud.

\item The cloud invokes $\Lambda.\proxyenc$ by taking as inputs ciphertext $C_i$ and $\rk_i$ $(1 \leq i\leq k)$. The ciphertext output by $\Lambda.\proxyenc$ is denoted by $C'_i$.
\end{itemize}

{\bf \noindent Evaluation Phase:}
\begin{itemize}

\item  The cloud invokes $\Lambda.\homevaluation$ by evaluating function $f$ with the inputs $C'_1,\cdots, C'_k$ and outputs a result $\outp$.
\end{itemize}

{\bf \noindent Decryption Phase:}

\begin{itemize}
\item  $k$ clients invoke $\Lambda.\thdec$ with their owned secret shares of private key $\ssk_1, \cdots, \ssk_k$ and $\outp$, so that they will obtain the result of $f(M_1, \cdots, M_k)$.
\end{itemize}

\subsection{A Homomorphic Threshold Proxy Re-encryption scheme}

To demonstrate the feasibility of the proposed encryption scheme, we present a such scheme, which allows the cloud to compute arbitrarily many multiplications with the ciphertexts under the target public/private keys.

\begin{itemize}

\item \setup($1^\ell,$):  Let $p$ be a $\ell-$bit prime, and let $G, G_T$ be two cyclic groups of order $p$. Let $e$ be a bilinear group, $e: G \times G \rightarrow G_T$. Let $g$ be the generator randomly selected from the group $G$, and $Z=e(g,g)$. The message space is $G_T$.

\item \keygen(i):  Client $i$ selects $\alpha_i$ from $\Z_p^*$ uniformly at random so that its public key will be $\pk_i = g^{\alpha_i}$ and the private key $\sk_i = \alpha_i$.

\item \thkeygen: The trusted third party selects $\alpha_{0}$ from $\Z_p^*$ uniformly at random. Let $\spk = g^{\alpha_0}$ and $\ssk=\{\ssk_1,\cdots,\ssk_k\}$, which is generated as follows: let $s(x) = \sum_{i=0}^{i=k}b_ix^i$, where $b_0= 1/\alpha_0$ and $b_i (1 \leq i\leq k)$ are selected from $\{1,\cdots, p-1\}$ uniformly at random. let $\ssk_i = ( i, s(i))$. Without loss of generality, we assume $k \geq 2$.

\item  \proxykeygen($\alpha_i,\spk$): Given client$i$'s private key $\sk_i = \alpha_i$ and the target public key $\spk = g^{\alpha_0}$ published by the third party,  client $i$ generates the proxy re-key $\rk_{i} = \spk^{1/\alpha_i}$.

\item \enc: Given message $M_i \in G_T$, client $i$ selects $r_i$ from $\Z_p^*$ uniformly at random and generates ciphertext $D_i = (C_{i1},C_{i2}$) as:
  \begin{equation*}
C_{i1} = \pk_i^{r_i} ,~~~~~ C_{i2} = Z^{r_i}M_i
  \end{equation*}

\item \proxyenc: Given the ciphertext $(C_{i1},C_{i2})$ under the public key $\pk_i$ and a re-encryption key $\rk_i = \spk^{1/\alpha_i}$, the proxy transforms the ciphertext to $(C'_{i1},C'_{i2})$ by:
  \begin{equation*}
C'_{i1} = e(C_{i1}, \rk_i) = e(C_{i1},\spk^{1/\alpha_i})  ,~~~~~
C'_{i2} = C_{i2}
  \end{equation*}

\item \homevaluation($(C'_{11},C'_{12}),(C'_{21},C'_{22}), \spk$): Given two ciphertext $(C'_{11},C'_{12}),(C'_{21},C'_{22})$ corresponding to the messages $M_1,M_2$ respectively, the ciphertext of the multiplication of $M_1\cdot M_2$ is $(C'_1, C'_2)$, where
  \begin{equation*}
  C'_1 =  C'_{11}\cdot C'_{21} ,~~~~~
  C'_2 =  C'_{12}\cdot C'_{22}
      \end{equation*}

\item \dec: Given the ciphertext $(C_{i1},C_{i2})$ under the public key $\pk_i$,  client $i$ decrypts it as follows:
  \begin{equation*}
 M_i = C_{i2} / e(C_{i1},g^{1/\alpha_i})  = Z^{r_i}M / Z^{r_i}
    \end{equation*}

\item \thdec:  Given the ciphertext $(C'_1, C'_2)$ under the target public key $\spk$, the decryption can be done with the cooperations of $k$ clients:

    for  client $i$, it computes
    $$ w_i = {C'}_1^{s(i)}, $$ and sends $w_i$ to all other clients, and then each client decrypts the ciphertext as
    $$ M = C'_2/ \prod_{i=1}^{i=k}(w_i)^{\lambda_{i}}, $$
    where
    $$\lambda_{i} = \prod_{j=1}^{j=k, j\neq i} \frac{j}{j-i}$$

\end{itemize}

\section{Conclusion}
\label{sec:conclusion}

\noindent
We initialize the study of multiparty cloud computing, where the cloud provides both storage service and computation service. The main goal is to enable many clients, by leveraging the cloud capability, to perform outsourced computation function in a secure and private manner. We propose the notion of homomorphic threshold proxy re-encryption scheme. Our ongoing work includes the construction of a fully (or somewhat) homomorphic threshold proxy re-encryption scheme and its security analysis.

\end{document}